\documentclass{aa}  

\usepackage{graphicx}
\usepackage{txfonts}

\usepackage{hyperref}

\usepackage{natbib}

\newcommand{\dgr}{^\circ}
\newcommand{\dd}{\mathrm{d}}
\newcommand{\domdt}{\dd\omega/\dd t}
\newcommand{\dadt}{\dd a/\dd t}

\begin{document} 

  \title{Rotation acceleration of asteroids (10115) 1992~SK, (1685)~Toro, and (1620)~Geographos due to the YORP effect}
  \titlerunning{Rotation acceleration of asteroids (10115) 1992~SK, (1685)~Toro, and (1620)~Geographos due to YORP}

  \author{J.~\v{D}urech         \inst{1}                        \and
          D.~Vokrouhlick\'y     \inst{1}        \and
          P.~Pravec             \inst{2}        \and
          Yu.~N.~Krugly         \inst{3}        \and
          M.-J.~Kim             \inst{4}        \and
          D.~Polishook          \inst{5}        \and
          V.~V.~Ayvazian                                \inst{6}                        \and
          T.~Bonev                                                      \inst{7}                        \and
          Y.-J.~Choi            \inst{4,8}      \and
          D.~G.~Datashvili                      \inst{6}                        \and
          Z.~Donchev                                            \inst{7}                        \and
          S.~A.~Ehgamberdiev            \inst{9}                        \and
          K.~Hornoch            \inst{2}        \and
                                        R.~Ya.~Inasaridze                 \inst{6,10}            \and
                                        G.~V.~Kapanadze                         \inst{6}                        \and
          D.-H.~Kim             \inst{4,11}             \and
          H.~Ku\v{c}\'akov\'a   \inst{2}        \and
          A.~V.~Kusakin                                 \inst{12}                       \and
          P.~Ku\v{s}nir\'ak     \inst{2}        \and
          H.-J.~Lee             \inst{4}        \and
          I.~E.~Molotov                                 \inst{13}                       \and
          H.-K.~Moon             \inst{4}       \and
          S.~S.~Mykhailova                       \inst{3}                        \and
          I.~V.~Nikolenko                               \inst{14}                       \and
          A.~Novichonok                                 \inst{15}                       \and
          J.~Oey                \inst{16}       \and
          Ch.~T.~Omarov                                 \inst{12}                       \and
          J.~T.~Pollock         \inst{17}       \and
          I.~V.~Reva                                            \inst{12}                       \and
          V~V.~Rumyantsev                               \inst{18}                       \and
          A.~A.~Zhornichenko                    \inst{15}                       
          }
 

  \institute{Astronomical Institute, Faculty of Mathematics and Physics, Charles University, V~Hole\v{s}ovi\v{c}k\'ach~2, 180\,00 Prague~8, Czech Republic\\
             \email{durech@sirrah.troja.mff.cuni.cz} \and
             Astronomical Institute, Academy of Sciences of the Czech Republic, Fri\v{c}ova~1, 251\,65 Ond\v{r}ejov, Czech Republic \and
             Institute of Astronomy of V.N. Karazin Kharkiv National University, 35 Sumska Str., Kharkiv 61022, Ukraine                                          \and
             Korea Astronomy and Space Science Institute, 776, Daedeokdae-ro, Yuseong-gu, Daejeon 34055, Republic of Korea          \and
             Faculty of Physics, Weizmann Institute of Science, 234 Herzl St., Rehovot 7610001, Israel                                                                                \and
             Kharadze Abastumani Astrophysical Observatory, Ilia State University, 3/5 K.Cholokoshvili Av., Tbilisi 0162, Georgia   \and
             Institute of Astronomy and NAO, Bulgarian Academy of Sciences, 72 Tsarigradsko Chaussee Blvd., BG-1784 Sofia, Bulgaria \and
                                                 University of Science and Technology, Korea                                                                                                                                                                                                                                                                                                               \and
                                                 Ulugh Beg Astronomical Institute, 33 Astronomicheskaya Str., Tashkent 100052, Uzbekistan                                                                                                                          \and
                                                 Samtskhe-Javakheti State University, 113 Rustaveli Str., Akhaltsikhe 0080, Georgia                                                                                                                                                       \and
                                                 Chungbuk National University, 1 Chungdae-ro, Seowon-gu, Cheongju, Chungbuk 28644, Korea                                                                                                                               \and
                                                 Fesenkov Astrophysical Institute, 23 Observatory, Almaty 050020, Kazakhstan                                                                                                                                                                               \and
                                                 Keldysh Institute of Applied Mathematics, RAS, 4 Miusskaya sq., Moscow 125047, Russia                                                                                                                                        \and
                                                 Institute of Astronomy, RAS, 48 Pyatnitskaya str., Moscow 119017, Russia                                                                                                                                                                                                \and
                                                 Petrozavodsk State University, 33 Lenin Str., Petrozavodsk 185910, Republic of Karelia, Russia                                                                                                 \and
                                                 Blue Mountains Observatory, 94 Rawson Pde. Leura, NSW 2780, Australia                                                  \and
                                         Physics and Astronomy Department, Appalachian State University, 525 Rivers St, Boone, NC 28608, USA                    \and
                                         Crimean Astrophysical Observatory, Nauchny, Crimea
             }

  \date{Received; accepted}

  \abstract
  {The rotation state of small asteroids is affected by the Yarkovsky–O'Keefe–Radzievskii–Paddack (YORP) effect, which is a net torque caused by solar radiation directly reflected and thermally reemitted from the surface. Due to this effect, the rotation period slowly changes, which can be most easily measured in light curves because the shift in the rotation phase accumulates over time quadratically.}
  {By new photometric observations of selected near-Earth asteroids, we want to enlarge the sample of asteroids with a detected YORP effect.}
  {We collected archived light curves and carried out new photometric observations for asteroids (10115) 1992~SK, (1620)~Geographos, and (1685)~Toro. We applied the method of light curve inversion to fit observations with a convex shape model. The YORP effect was modeled as a linear change of the rotation frequency $\upsilon \equiv \domdt$ and optimized together with other spin and shape parameters.}
  {We detected the acceleration $\upsilon = (8.3 \pm 0.6) \times 10^{-8}\,\text{rad}\,\text{d}^{-2}$ of the rotation for asteroid (10115) 1992~SK.  This observed value agrees well with the theoretical value of YORP-induced spin-up computed for our shape and spin model. For (1685)~Toro, we obtained $\upsilon = (3.3 \pm 0.3) \times 10^{-9}\,\text{rad}\,\text{d}^{-2}$, which confirms an earlier tentative YORP detection. For (1620)~Geographos, we confirmed the previously detected YORP acceleration and derived an updated value of $\upsilon$ with a smaller uncertainty. We also included the effect of solar precession into our inversion algorithm, and we show that there are hints of this effect in Geographos' data.}
  {The detected change of the spin rate of (10115) 1992~SK has increased the total number of asteroids with YORP detection to ten. In all ten cases, the $\domdt$ value is positive, so the rotation of these asteroids is accelerated. It is unlikely to be just a statistical fluke, but it is probably a real feature that needs to be explained.}


  \maketitle

  \section{Introduction}
  
    The importance of nongravitational radiation forces for spin evolution of asteroids was fully recognized by \cite{rub2000}, who coined the term Yarkovsky–O'Keefe–Radzievskii–Paddack (YORP) effect for solar radiation-induced torque that affects the rotation state of asteroids. The YORP effect is important for the evolution of the asteroid population, as asteroids can be accelerated to the rotation break limit, they can shed mass and create asteroid pairs or binaries. It also affects the distribution of rotation rates and obliquities in general. For details and further references, see the review of \cite{vetal2015}. 
    
    Research on theoretical aspects of YORP \citep[][for example]{Bre.ea:07, Nes.Vok:07, Bre.Vok:11, rg2012} went hand in hand with efforts to detect this effect directly as a change in the rotation period \citep{ketal2007, Low.ea:07, Tay.ea:07}. With the new YORP detections presented in this paper, the number of asteroids known to change their rotation period due to YORP has grown to ten, all of them being small near-Earth asteroids because the magnitude of the YORP  effect is inversely proportional to the squared heliocentric distance and squared size. After five asteroids listed in the review of \cite{vetal2015}, there were four more: (161989)~Cacus and (1865)~Toro (only tentative detection, now confirmed in our paper) in \cite{Dur.ea:18a}, (101955)~Bennu \citep{Nol.ea:19b, hetal2019}, and (68346) 2001~KZ66 \citep{Zeg.ea:21}. To these nine, we added the tenth detection of the rotation period change for asteroid (10115) 1992~SK. This, as all previous detections, also has a positive sign of $\domdt$, so its rotation is accelerated.
    
  \section{Reconstructing the spin state and shape from light curves}
  \label{sec:LC_inversion}
  
    To detect changes in the rotation periods that are too small to be found directly \citep[as in the case of the asteroid YORP,][]{Low.ea:07}, it is necessary to look for shifts in the rotation phase. Contrary to the rotation period, which evolves linearly when affected by YORP, the rotation phase drift accumulates over time and increases quadratically with time. We used the same approach as in \cite{ketal2007} or \cite{Dur.ea:18b} -- the change in the rotation rate $\omega$ is described by a free parameter $\upsilon \equiv \domdt$ that is optimized during the light curve  inversion together with the shape and spin parameters. If a nonzero $\upsilon$ provides a significantly better fit than $\upsilon = 0$, we interpret it as detecting rotation acceleration or deceleration. The next step is to show that this observed value of $\upsilon$ is consistent with the YORP value predicted theoretically from the known shape, size, and spin of the asteroid (see Sect.~\ref{sec:theoretical_YORP}). The light curve inversion method iteratively converges to best-fit parameters that minimize the difference between the observed and modeled light curves; this difference is measured by the standard $\chi^2$. To realistically estimate uncertainties of photometric data, we fit each light curve by a Fourier series of maximum order determined by an F-test \citep{Mag.ea:96}. The root-mean-square residual was used as the uncertainty of individual light curve points. In other words, light curves were weighted according to their precision.    
    
    In the following subsections, we present spin parameters (listed also in Table~\ref{tab:parameters}) obtained as best-fit parameters by light curve inversion \citep{Kaa.ea:01}. Their uncertainties were estimated by a bootstrap method. For each asteroid we analyzed, we created 10,000 bootstrapped light curve data sets by randomly selecting a new set of light curves with a random resampling of light curve points. For each new data set, we repeated the inversion and obtained spin parameters. From the distribution of these parameters, we estimated their uncertainties. We also estimated the uncertainty of the $\upsilon$ parameter by varying it around its best value and looking at the increase in $\chi^2$. Error intervals provided by this approach are slightly larger than those determined by bootstrap for (10115) and Geographos and twice as large for Toro (see the Appendix for details).
    
    \subsection{(10115) 1992 SK}
  
      The first light curves of this asteroid were observed in 1999, and they were used together with radar delay-Doppler observations for the shape reconstruction by \cite{Bus.ea:06}. Other photometry comes from 2006 \citep{Pol:12} and 2013 \citep{War:14e}. We observed this object during two apparitions in 2017 and 2020. Using these calibrated observations and also those from February to March 1999, we determined its mean absolute magnitude $H = 17.31 \pm 0.18$\,mag, assuming the slope parameter $G = 0.24 \pm 0.11$  \citep[which is the $1\sigma$ range of G values for S type asteroids from][]{War.ea:09}. The color index in the Johnson-Cousins system is $V - R = 0.458 \pm 0.013$\,mag, consistent with its Sq/S spectral classification of \cite{Tho.ea:14} and \cite{Bi.ea:19}. All available light curves are listed in Table~\ref{tab:aspect_10115}.
      
\begin{figure}[t]
        \begin{center}
          \includegraphics[width=\columnwidth]{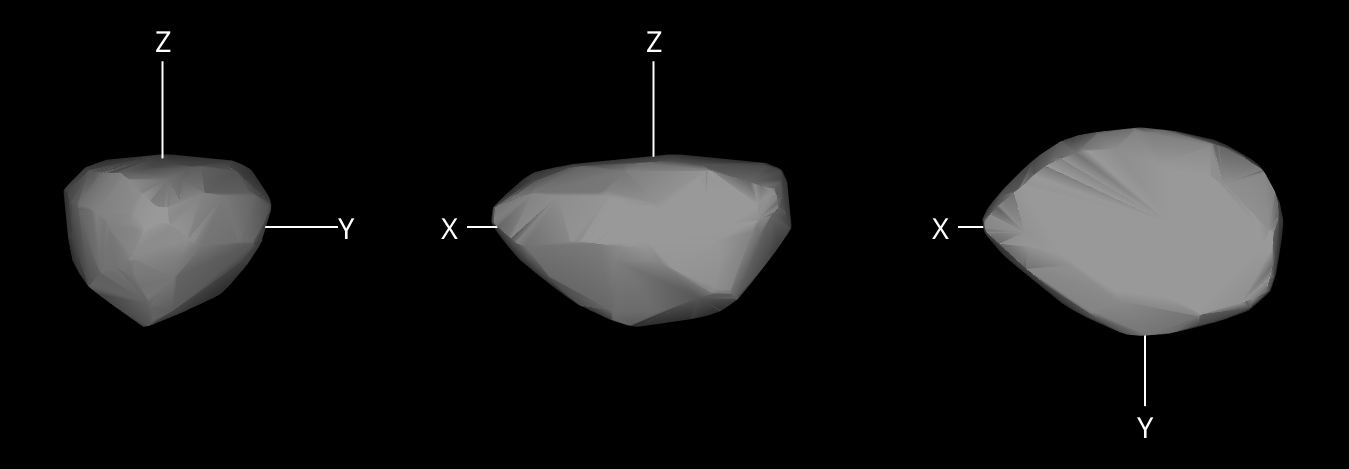}
          \caption{\label{fig_10115_shape}
          Shape model of asteroid (10115) 1992~SK shown from equatorial level (left and center, $90\degr$ apart) and pole-on (right).}
        \end{center}
      \end{figure}

\begin{figure*}[t]
        \begin{center}
          \includegraphics{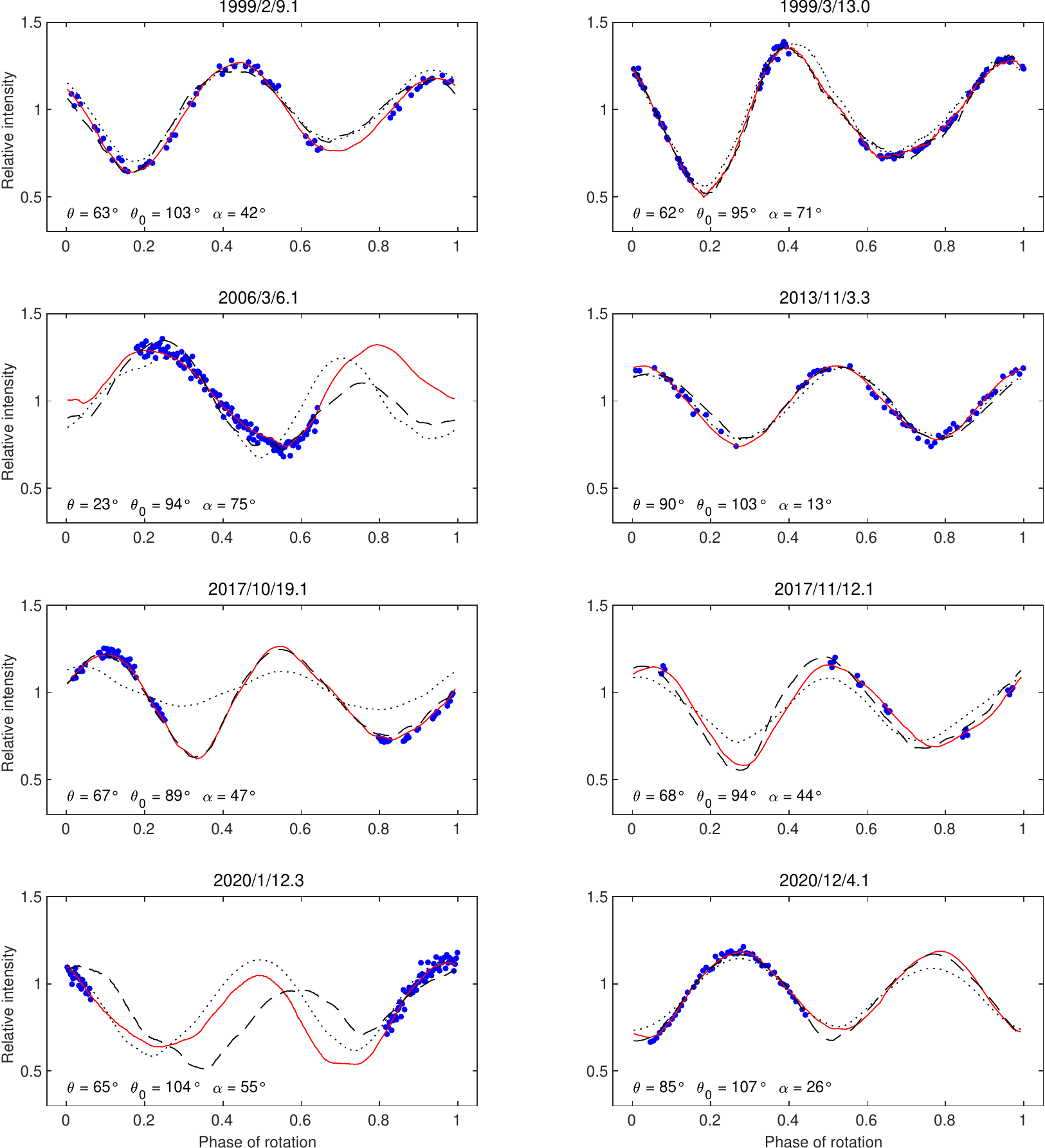}
          \caption{\label{fig_10115_lc}
          Example light curves (blue dots) of (10115) 1992~SK shown with the synthetic light curves produced by the best YORP model (red curves) and the best constant-period model (black dashed curves). We also show synthetic light curves generated by the shape and spin model of \cite{Bus.ea:06} (dotted black curves). The latter were manually shifted in phase to give the best fit with observed data, but still, their amplitude is not consistent with 2017 observations. The geometry of the observation is described by the aspect angle $\theta$, the solar aspect angle $\theta_0$, and the solar phase angle $\alpha$.}
        \end{center}
      \end{figure*}

\begin{figure}[t]
        \begin{center}
          \includegraphics[width=\columnwidth, trim=1cm 4cm 1cm 4.7cm, clip]{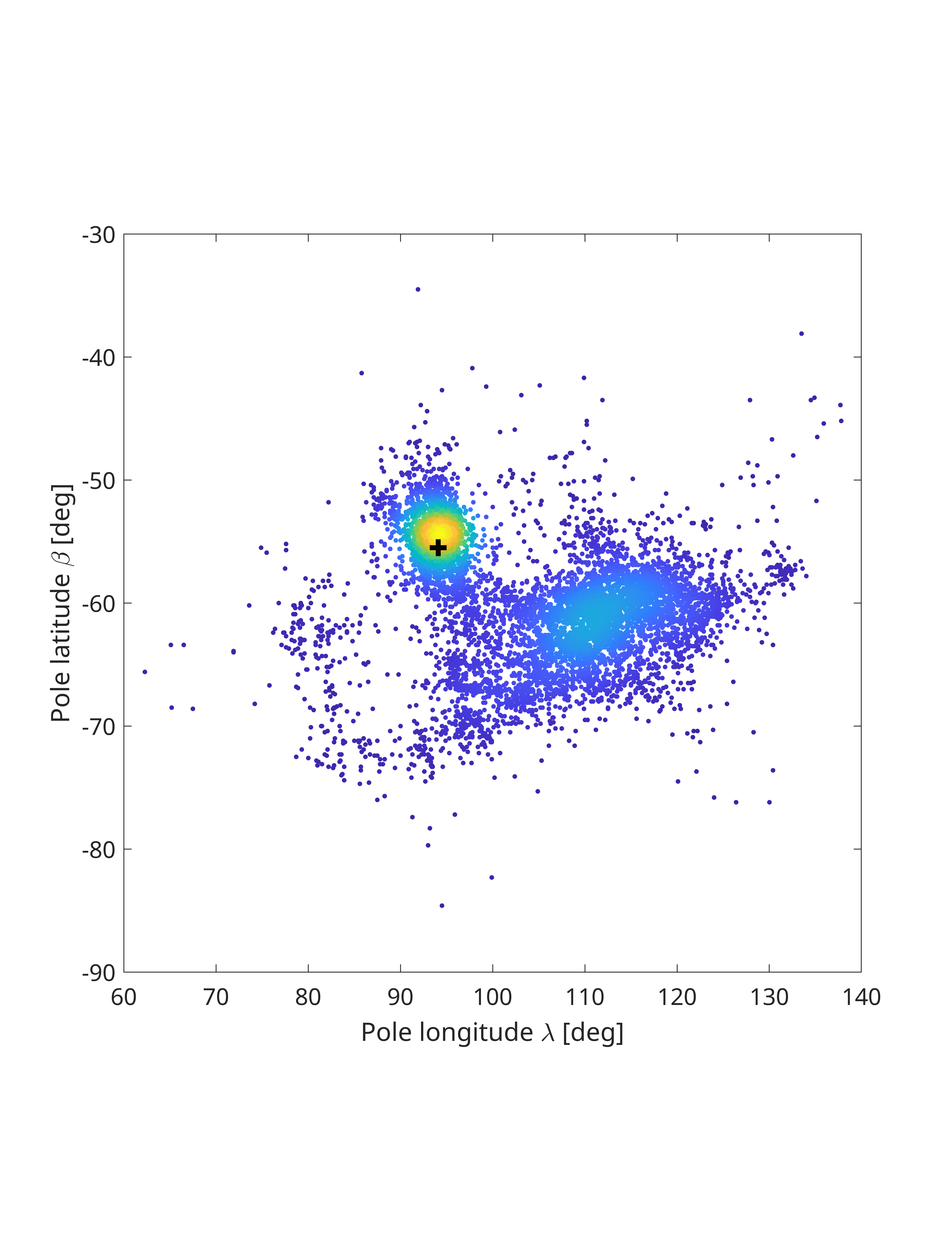}
          \caption{\label{fig_10115_pole}
          Direction of the spin axis in ecliptic coordinates $(\lambda, \beta)$ for 10,000 bootstrap models of (10115) 1992~SK. The density of points is color-coded, with the red color corresponding to the highest density. The black cross at $(95^\circ, -54^\circ)$ marks the spin-axis direction of the nominal model derived from the original light curve data set.}
        \end{center}
      \end{figure}

      The light curve inversion provided unambiguous results; the detection of the period change is robust, that is, a model with a constant rotation period provides a significantly worse fit to light curves than the YORP model. The best-fit model has a pole direction in ecliptic coordinates $(94^\circ, -56^\circ)$, the rotation period $P = 7.320232 \pm 0.000010$\,h (for JD 2451192.0), and the YORP value $\upsilon = (8.3 \pm 0.6) \times 10^{-8}\,\text{rad}\,\text{d}^{-2}$ ($1\sigma$ errors). The shape model is shown in Fig.~\ref{fig_10115_shape} and the agreement between synthetic light curves produced by this shape and real observations is demonstrated in Fig.~\ref{fig_10115_lc}. The distribution of bootstrap results for the spin axis direction was bi-modal with $\lambda$ in the range 90--120$^\circ$ and $\beta$ between $-65$ and $-45^\circ$ (see Fig.~\ref{fig_10115_pole}). Formal $1\sigma$ uncertainties computed as standard deviations were $\pm 10^\circ$ for $\lambda$ and $\pm 5^\circ$ for $\beta$. The bi-modality is caused by the random selection of light curves in bootstrap and the importance of some light curves -- the cloud of points around the pole direction $(110^\circ, -60^\circ)$ are mainly those solutions that do not have the light curve from 2020 December~4 in the bootstrap input data. Random resampling also causes the best-fit pole direction based on the original light curve data set to not be exactly at the position where the density of bootstrap solutions is the highest.

      The radar-based model of \cite{Bus.ea:06} has a similar shape as our convex model and a similar rotation period of $7.3182 \pm 0.0003$\,h (although the periods are different when measured by their $3\sigma$ uncertainty intervals), but its spin axis orientation $(99^\circ, -3^\circ)$ is significantly different from our value. Their model is not consistent with new light curves from 2017 -- the light curve amplitudes are not correctly reproduced (see Fig.~\ref{fig_10115_lc}). 

    \subsection{(1620) Geographos}
    \label{sec:Geographos}
  
      There are a lot of photometric observations of Geographos going back to 1969. Geographos was also observed by radar and a shape model was reconstructed \citep{Hud.Ost:99} and thermohpysical analysis was performed by \cite{Roz.Gre:14}. YORP-induced acceleration of its rotation was detected by \cite{Dur.ea:08b} by the inversion of light curves from 1969 to 2008. The rotation parameters were determined to ($3\sigma$ uncertainties): $\lambda = 58 \pm 6\dgr$, $\beta = -49 \pm 7\dgr$, $P = 5.223336 \pm 0.000002$\,h, and $\upsilon = (1.15 \pm 0.15) \times 10^{-8}\,\text{rad\,d}^{-2}$.
      
      We complemented the previous data set with new photometry from 2008, 2011, 2012, 2015, and 2019 (see Table~\ref{tab:aspect_1620}) and updated the spin parameters to the following new values ($1\sigma$ uncertainties): $\upsilon = (1.14 \pm 0.03) \times 10^{-8}\,\text{rad}\,\text{d}^{-2}$ with period $P = 5.2233360 \pm 0.0000006\,\text{h}$ (for JD 2440229.0). The pole direction $\lambda = 56.7 \pm 0.7^\circ$, $\beta = -51.2 \pm 1.1^\circ$ in ecliptic coordinates. The shape model is shown in Fig.~\ref{fig_1620_shape}.

    \subsection{(1685) Toro}

      A tentative YORP detection by \cite{Dur.ea:18a} was based on a data set from apparitions in 1972 to 2016. To this, we added new observations from 2018, 2020, and 2021 (see Table~\ref{tab:aspect_1685}). We determined Toro's mean absolute magnitude to $H = 14.48 \pm 0.13$\,mag, assuming the slope parameter $G = 0.24 \pm 0.11$ \citep{War.ea:09}. We measured the color index in the Johnson-Cousins system as $V - R = 0.462 \pm 0.010$\,mag, which is consistent with its Sq/S spectral classification of \cite{Tho.ea:14} and \cite{Bi.ea:19}.
    
\begin{figure*}[t]
        \begin{center}
          \includegraphics{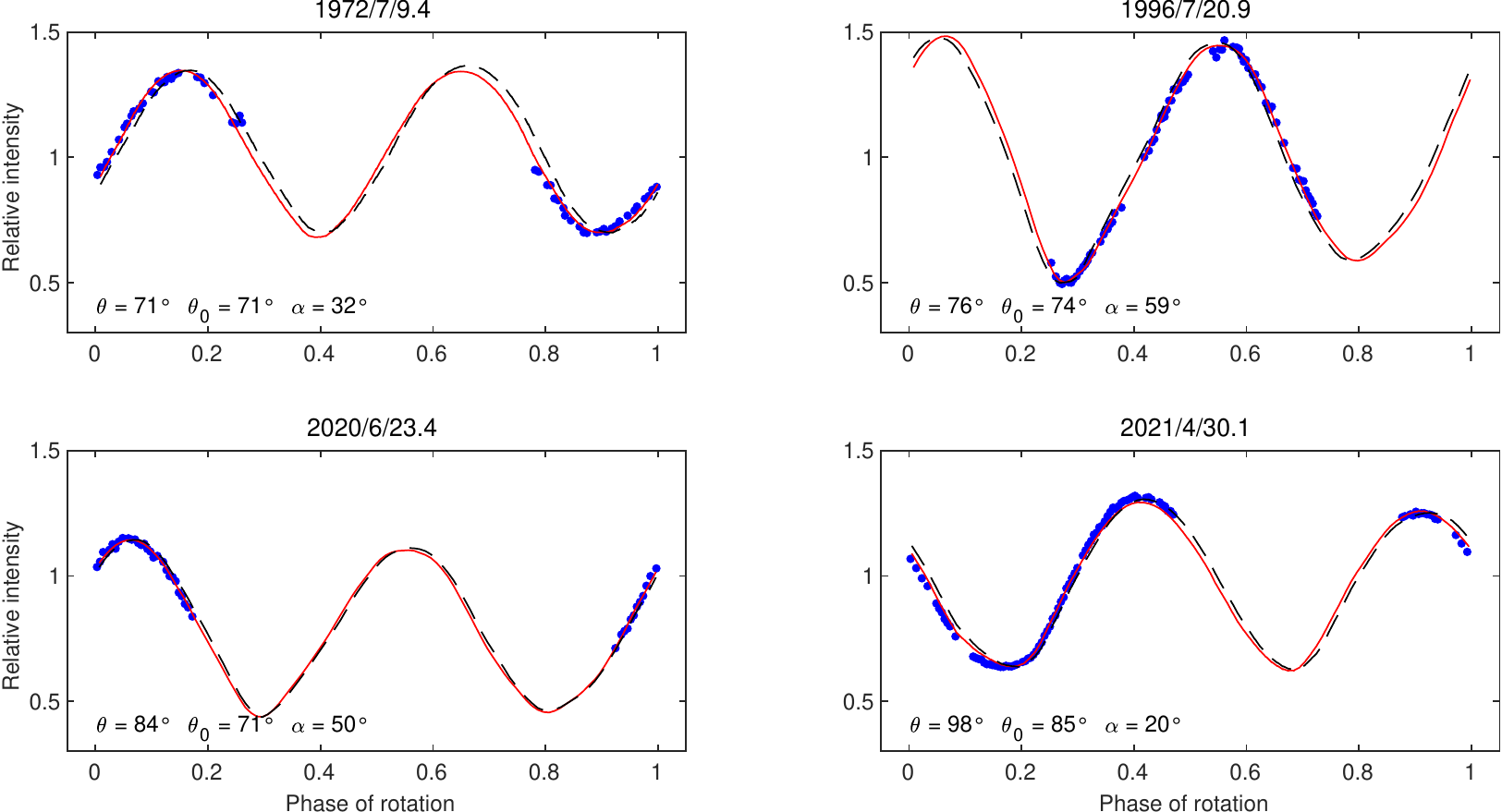}
          \caption{\label{fig_1685_lc}
          Example light curves (blue dots) of (1685)~Toro shown with the synthetic light curves produced by the best YORP model (red curves) and the best constant-period model (black dashed curves). The geometry of observation is described by the aspect angle $\theta$, the solar aspect angle $\theta_0$, and the solar phase angle $\alpha$.}
        \end{center}
      \end{figure*}

      Previous values ($3\sigma$ errors) published by \cite{Dur.ea:18a} were $P = 10.19782 \pm 0.00003\,$h, $(\lambda, \beta) = (71 \pm 10\dgr, -69 \pm 5\dgr)$, and $\upsilon = 3.0 \times 10^{-9}\text{rad\,d}^{-2}$. The updated values ($1\sigma$ errors) are as follows: $\upsilon = (3.3 \pm 0.3) \times 10^{-9}\,\text{rad}\,\text{d}^{-2}$, $P = 10.197826 \pm 0.000002\,\text{h}$ (for JD 2441507.0). The pole direction is $(75 \pm 3^\circ, -69 \pm 1^\circ)$. The shape model is shown in Fig.~\ref{fig_1685_shape}. Although the model with YORP is significantly better than a constant-period model statistically, the difference between the synthetic light curves they produce is so small that they look almost the same when plotted on top of each other. In Fig.~\ref{fig_1685_lc}, we show four light curves for which the difference between the two models is the largest.

  \begin{table*}[t]
  \setlength{\tabcolsep}{3pt}
    \caption{\label{tab:parameters}     
        Parameters derived from photometric data: Spin axis direction in ecliptic longitude $\lambda_\text{p}$   and latitude $\beta_\text{p}$, the sidereal rotation period $P$ at epoch JD$_0$, the YORP parameter $\upsilon$, the absolute magnitude $H$, and the color index $V - R$. The volume-equivalent diameters $D$ were taken from \cite{Hud.Ost:99} for Geographos, from \cite{Dur.ea:18a} for Toro, and from \cite{Bus.ea:06} for 1999~SK.}
        \centering
        \begin{tabular}{r@{\hspace{3pt}}lccrccccc}
            \hline \hline\\[-3mm]
            \multicolumn{2}{c}{Asteroid}        & $\lambda_\text{p}$     & $\beta_\text{p}$      & \multicolumn{1}{c}{$P$}  & JD$_0$       & $\upsilon$                       & $D$     & $H$                & $V - R$   \\
                        &                 & [deg]                    & [deg]                         & \multicolumn{1}{c}{[h]}  &              & [$\times 10^{-8}\,\text{rad}\,\text{d}^{-2}$]       &    [km]   & [mag]            & [mag] \\
            \hline\\[-3mm]                
            (1620)      & Geographos      & $56.7 \pm 0.7$       & $-51.2 \pm 1.1$  & $5.2233360$  & 2440229.0 & $1.14 \pm 0.03$ & $2.56 \pm 0.15$      &                  &        \\
                        &                 &                      &                  & $ \pm 0.0000006$   &           &                                  &         &                  &        \\       
            (1685)      & Toro            & $75 \pm 3$           & $-69 \pm 1$      & $10.197826$    & 2441507.0 & $0.33 \pm 0.03$ & $3.5^{+0.3}_{-0.4}$& $14.48 \pm 0.13$ & $0.462 \pm 0.010$ \\
                        &                 &                      &                  & $ \pm 0.000002$    &           &                                  &         &                  &        \\       
            (10115)     & 1992 SK         & $94 \pm 10 $         & $-56 \pm 5$      & $7.320232$   & 2451192.0 & $8.3 \pm 0.6$   & $1.0 \pm 0.2$        & $17.31 \pm 0.18$ & $0.458 \pm 0.013$  \\
                        &                 &                      &                  & $ \pm 0.000010$    &           &                                  &         &                  &        \\       
            \hline
        \hline
        \end{tabular}
    \end{table*}
    
  \section{Solar torque precession}
  
    The detection of small secular changes on the order of $\dd P / \dd t \sim 10^{-11}$ to $10^{-9}$ in the rotation period is possible due to the long time span of photometric observations. In the case of Geographos, the data cover 50 years, and the large amplitude of its light curves enabled us to determine the rotation period and its secular change very precisely. For the same reasons, the spin axis' direction was determined with an exquisite precision of about $1\dgr$. 
    
    So far, all light curve inversion models have assumed that the rotation axis is fixed in the inertial frame, which was a valid assumption for asteroids in principal axis rotation. Apart from changing the angular frequency, the YORP effect also causes a secular evolution of the spin axis obliquity, but this effect is so tiny that it is unobservable with current data sets. For example, the theoretical change in Geographos' obliquity is smaller than one arcminute in 50~yr. However, there is another effect -- a regular precession due to the solar gravitation torque -- that inevitably affects the direction of the spin axis of all asteroids.
    
    Solar gravitation torque acts on a rotating body and causes a secular precession of its rotation axis around the normal to its orbital plane. The precession constant $\alpha$ describes the angular velocity of the spin axis at the limit of zero obliquity and can be expressed as
    \begin{equation}
      \label{eq:alpha}
      \alpha = \frac{3 n^2 \Delta}{2 \omega \eta^3}\,,
    \end{equation}
    where $\eta$ is determined from eccentricity $e$ as $\eta = \sqrt{1 - e^2}$, $\Delta$ is the dynamical ellipticity computed from the principal values of the inertia tensor $A, B, C$  as $1 - 0.5 (A + B) / C$, $\omega$ is the angular rotational velocity $\omega = 2\pi/P$, and $n$ is the orbital mean motion \cite[e.g,][chapter 4]{Ber.ea:03}. If we substitute the values $\Delta = 0.29 \pm 0.02$ for Toro and $\Delta = 0.407 \pm 0.006$ for Geographos (uncertainties estimated by bootstrap), we get $\alpha = (356 \pm 27)\arcsec / \text{yr}$ and $(289 \pm 4)\arcsec / \text{yr}$, respectively. This formally accumulates to $\sim 4\dgr$ over 50 years. The motion of the rotation axis on the precession cone with obliquity $\epsilon$ has angular velocity $\Delta\varphi / \Delta t = \alpha \, \cos\epsilon$ and mainly affects the ecliptic longitude $\lambda$ of the spin axis for orbits of a moderately small inclination (see the next subsection). Even though the $\cos\epsilon$ factor slightly decreases the above-estimated effect, the formal uncertainty of the pole determination for Geographos is so small that precession should have a measurable effect on its photometric data.
      
    \subsection{Model of solar precession}
    
      We included solar precession in the light curve inversion algorithm to see if it somehow affects our results. The evolution of the unit spin vector $\vec{s}(t)$ over small time interval $\Delta t$ is described as $\vec{s}(t + \Delta t) = \vec{s}(t) + \Delta\vec{s}(t)$, where the incremental change $\Delta\vec{s} = - (\vec{N} \times \vec{s})\, \Delta\varphi$. The normal to the orbital plane $\vec{N}$ is defined by means of the inclination $i$ of the orbital plane to the ecliptic and the longitude of the ascending node $\Omega$ as: $\vec{N} = (\sin i \sin\Omega; -\sin i \cos\Omega; \cos i)^\text{T}$. We assume that $i$ is constant and that $\Omega$ evolves linearly in time as $\Omega = \Omega_0 + \dot{\Omega} t$. We took the values of $\dot{\Omega}$ from the NEODyS page\footnote{\url{https://newton.spacedys.com/neodys/}}, but the results with a constant $\Omega$ were practically the same because $\dot{\Omega}$ is an order of magnitude smaller than $\alpha$. The real angular shift $\Delta\varphi$ of the spin vector is $\Delta\varphi = \alpha \cos\epsilon\,\Delta t = \alpha \, (\vec{s} \cdot \vec{N}) \, \Delta t$.
      
      The initial (at the epoch of the first light curve) orientation of the spin vector $\vec{s}_0$ is described by the ecliptic coordinates $\lambda_0$ and $\beta_0$, both being free parameters of optimization. Contrary to the standard light curve inversion, the pole direction is not fixed in space but evolves due to precession according to the equations given above. 
      
\begin{figure}[t]
        \begin{center}
          \includegraphics[width=\columnwidth]{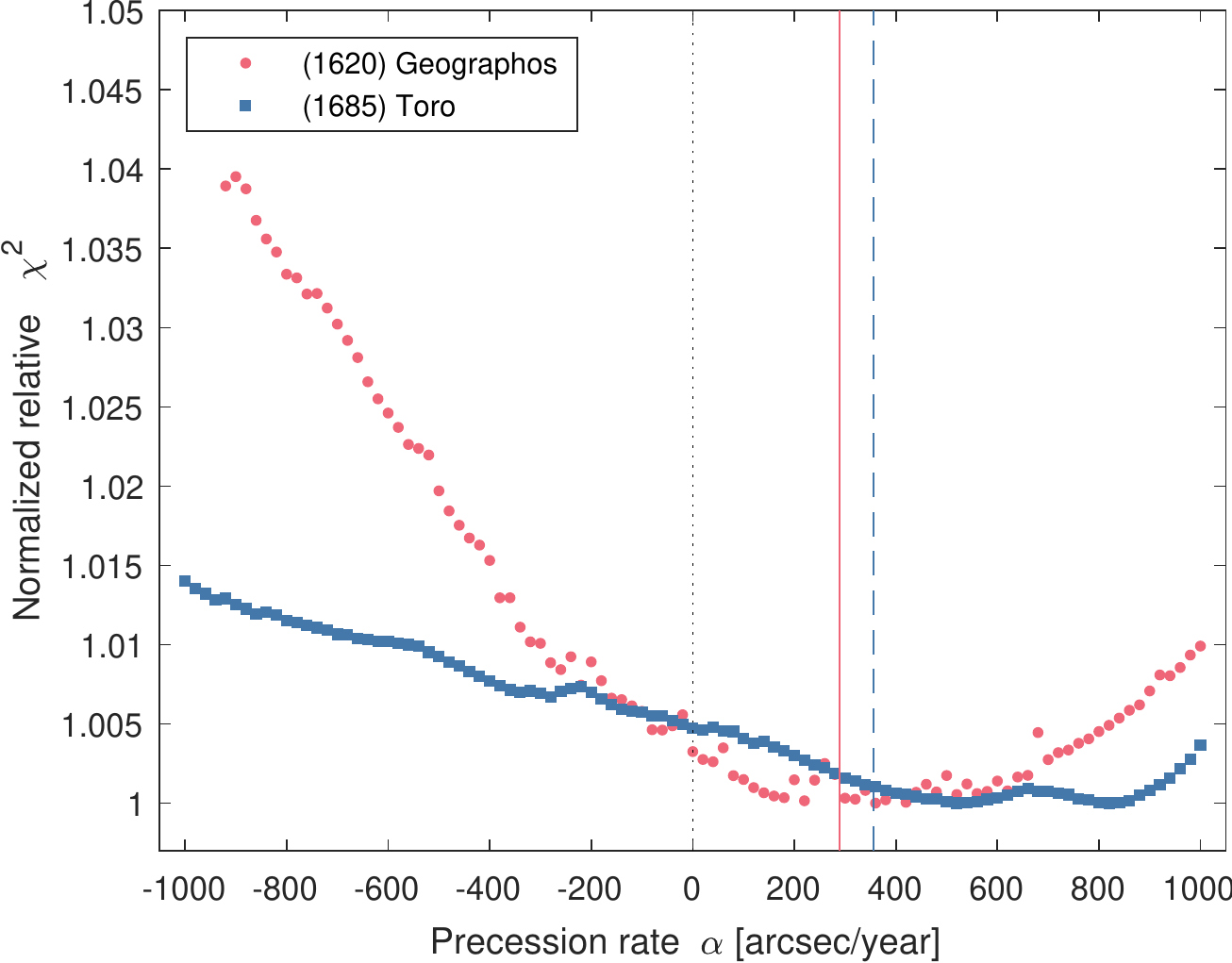}
          \caption{\label{fig:1620_1685_prec}
          Normalized relative $\chi^2$ versus precession rate $\alpha$ for (1620)~Geographos and (1685)~Toro. The nominal values of $\alpha = 289\arcsec/\text{yr}$ for Geographos (solid) and $\alpha = 356\arcsec/\text{yr}$ for Toro (dashed) are marked with vertical lines. The iterative convergence of the inversion algorithm is not ideal and causes some scatter of points mainly on Geographos' curve.}
        \end{center}
      \end{figure}

      For Geographos, the precession constant is $\alpha = 289\,\arcsec/\text{yr}$, which implies the accumulated shift $\Delta\varphi = -3.54\dgr$ over 50 years. The corresponding change of ecliptic coordinates of the spin axis is $\Delta\lambda = 2.44\dgr$ and $\Delta\beta = 0.13\dgr$. This expected shift of the ecliptic longitude $\lambda$ is larger than its formal uncertainty of $0.7\dgr$ (Sect.~\ref{sec:Geographos}), so precession should have a measurable effect on Geographos' light curves. Indeed, including the evolution of $\vec{s}$ into the inversion has a small, yet statistically significant effect on the goodness of the fit. We changed the precession constant $\alpha$ on the interval from $-1000$ to $1000\arcsec/\text{yr}$ with a step of $20 \arcsec/\text{yr}$, and for each value, we repeated the light curve inversion, that is to say we optimized all shape and spin parameters, also including the YORP parameter $\upsilon$. The results are shown in Fig.~\ref{fig:1620_1685_prec}, where $\chi^2$ values are plotted for different values of $\alpha$. Values of $\chi^2$ were rescaled such that the minimum value was equal to one. We can see a clear asymmetry with respect to zero precession -- the lowest $\chi^2$ is obtained for $\alpha$ around 200--600\,\arcsec/\text{yr}. Without any precession ($\alpha = 0$), the $\chi^2$ is about 0.5\% higher than the minimum value, which is just a small increase, but still statistically significant given the number of data points (8852 in total). Moreover, the best fit is obtained for values of $\alpha$ that agree with the theoretical value of $\sim 300\arcsec/\text{yr}$, so we interpret this as a detection of the precession of the Geographos' rotation axis due to the solar torque.
    
      For Toro, the length of the time interval covered by observations (48 years, 6995 data points) is about the same as for Geographos; the precession constant $\alpha$ is similar, so the precession evolution is about the same: $\Delta\varphi = -4.36\dgr$, $\Delta\lambda = 3.64\dgr$, and $\Delta\beta = -0.66\dgr$. However, the spin axis direction is not so tightly constrained, and including the parameter $\alpha$ into the model has a much smaller effect than in the case of Geographos. The dependence of $\chi^2$ on $\alpha$ is much weaker (Fig.~\ref{fig:1620_1685_prec}). Although the $\chi^2$ versus $\alpha$ curve is also not symmetric around zero and positive values of $\alpha$ are preferred (the best fit is obtained for $\alpha \sim 400$--$900\,\arcsec/\text{yr}$), it is not robust with respect to the data set -- that is, excluding some light curves from the data set or changing their formal errors has a strong effect on the shape of the $\alpha$ versus $\chi^2$ dependence. We repeated the $\alpha$ scan with bootstrapped light curve samples and confirmed that for Geographos, the results were much more stable than for Toro. From the sample of one hundred bootstrap repetitions, the value of $\alpha$ for which $\chi^2$ was minimal was $(330 \pm 360)\,\arcsec/\text{yr}$ for Geographos, while for Toro it was $(-110 \pm 620)\,\arcsec/\text{yr}$. 
      
      The dynamical ellipticity of asteroid 1992~SK is $\Delta = 0.29$, which leads to theoretical $\alpha = 283\,\arcsec/\text{yr}$. Accumulated over 22 years of observations, $\Delta\varphi = -1.56\dgr$, which is too small given the large uncertainties of the pole direction -- thus this effect is not detectable with the current data set.
      
      Geographos and Toro are almost ideal candidates for the search of the effect of precession on photometric data -- they are near-Earth asteroids (large $n$), elongated (large $\Delta$), and not rotating too quickly (low $\omega$). It is not likely that many asteroids will have $\alpha$ much larger than a couple of hundreds of arcsec per year. From the modeling point of view, precession will continue to be negligible for most of the asteroids. However, for slowly rotating elongated near-Earth asteroids with suitable obliquity and observations covering many decades, this effect will become measurable and should be included in the modeling. In principle, the precession rate $\alpha$ can be treated as another free parameter of the light curve fitting procedure, so $\Delta$ can be determined (Eq.~\ref{eq:alpha}) from observations independently of the shape model. Its value may then be compared with that computed from the shape model assuming a uniform density distribution. 
      
\section{Comparison of the detected $\upsilon$ values with the theoretical model}
\label{sec:theoretical_YORP}

A secular change in the rotation rate of the three objects presented in Sec.~\ref{sec:LC_inversion}
has been obtained using a fully empirical approach, namely by adjusting the rate
factor $\upsilon=\domdt$ to match the observations. In order to physically
interpret $\upsilon$, it is now important to compare it with a model prediction.
We postulate that the YORP effect is the underlying mechanism that makes the rotation
rate of (10115) 1992~SK, (1685) Toro, and (1620) Geographos accelerate, and we therefore use
a model description of YORP to predict the expected level of $\upsilon_{\rm mod}$.
A comparison between $\upsilon$ and $\upsilon_{\rm mod}$ helps to verify, or reject,
our assumption, and eventually argue about parameters on which $\upsilon_{\rm mod}$
depends. A well-known issue with the YORP model exists because (i) some
of its free parameters are apparent from the observations, such as the rotation state
or size, (ii) some may be plausibly expected and described by a reliable statistical
distribution, such as the bulk density, but (iii) some others cannot be easily
described or determined. This last class depends on the small-scale irregularities of
the asteroid shape. Yet, numerical tests confirmed that $\upsilon_{\rm mod}$ may
in some circumstances strongly depend on this last class of parameters 
\citep[e.g.,][]{s2009,rg2012,rg2013,gk2012}.

With this caveat in mind, we used a fully thermophysical model of \citet{Cap.Vok:04}, see also
\citet{Cap.Vok:05}, to evaluate the surface temperature on a body, whose shape is
represented with a general polyhedron. The number of surface facets $N$, provided by inversion
of the photometric observations, is typically a couple thousand. As a rule of thumb, this
allows one to resolve surface features of a characteristic scale $\propto D\,\sqrt{\pi/N}$,
that is, some $\simeq 40$ meters at best for kilometer-sized asteroids. Our approach treats
each of the surface facets individually and solves the one-dimensional problem of
heat conduction below the surface. A fixed configuration of the heliocentric orbit and
rotation pole of the body is assumed. Boundary conditions express (i) energy conservation
at the surface (with solar radiation as an external source) and (ii) a constant temperature
at great depth (implying an isothermal core of the body). Given a plausible range of surface
thermal conductivity values and typical rotation rates of asteroids, the penetration depth
of a thermal wave is in the centimeter to meter range. With the coarse resolution of the
shape model, the one-dimensional treatment of the problem is justified. The model, in principle, accounts
for mutual shadowing of surface units, but neglects their mutual irradiation
\citep[e.g.,][]{rg2012} and possible thermal communication via conduction \citep[which would
require a resolution model at the level of penetration depth of the thermal wave, e.g.,][]{gk2012}.
We also assume simple Lambertian thermal emission from the surface, neglecting thermal
beaming related again to the small-scale surface roughness \citep[e.g.,][]{rg2013}.
All these complications must be accounted for with an empirical correction. Once the surface
temperature was determined for any of the surface facets, and at any moment during the revolution
about the Sun, the thermal recoil force and torque were determined (in the torque part, we also
added effects of the directly reflected sunlight in the optical waveband). The force part may
help to determine the orbit-averaged change $\dadt$ of the heliocentric semimajor axis (the
Yarkovsky effect), and the torque provides the orbit-averaged change $\upsilon=\domdt$ of
the rotation rate (the YORP effect). Technical details may be found in the abovementioned
papers by \citet{Cap.Vok:04} and \citet{Cap.Vok:05}.

In our simulations, we tested surface thermal conductivity values in
the range of $40$ to nearly $1200$ J~m$^{-1}$~K$^{-1}$~s$^{-1/2}$ which are appropriate
for kilometer-sized near-Earth asteroids \citep[e.g.,][]{delbo2015}.
All three asteroids discussed in Sec.~\ref{sec:LC_inversion} are spectrally S-type. This
justifies our nominal choice of the bulk density $\rho=2.5$ g~cm$^{-3}$, though
slightly larger values have been determined for larger asteroids, while
possibly smaller values for small near-Earth objects \citep[see,
e.g.,][]{schetal2015}. A realistic uncertainty in the adopted bulk density
may, therefore, be $\simeq 0.7$ g~cm$^{-3}$. Parameters of the rotation state,
namely the rotation period and pole, were taken from our solution in Sec.~\ref{sec:LC_inversion}, and
the parameters of heliocentric orbits from standard databases (such as JPL or
AstDyS).

\subsection{(10115) 1992~SK}
Using the physical parameters mentioned above and scaling our convex-shape
model such that it has a volume equivalent to a sphere of diameter $1$~km \citep{Bus.ea:06}, we
obtained $\upsilon_{\rm mod} = 19.3\times 10^{-8}$ rad~d$^{-2}$ by YORP. As
already found by \citet{Cap.Vok:04}, and later verified by both numerical and
analytical means \citep[e.g.,][and references therein]{vetal2015}, the
predicted $\upsilon_{\rm mod}$
does not depend on the surface thermal conductivity in the simple one-dimensional
heat conduction approach used. So our predicted value potentially rescales only
with adjustments to the size $D$ and bulk density $\rho$ with $\upsilon_{\rm mod}
\propto 1/(\rho D^2)$. Nonetheless, the difference between our base value of 
$\upsilon_{\rm mod}$ and the observed $\upsilon = (8.3\pm 0.6)\times 10^{-8}$
rad~d$^{-2}$ is comfortably small, in fact smaller than in some other cases of
asteroids with YORP detections. Additionally, it
can be made closer to the observed value by (i) a plausible small increase in the bulk
density and/or size, or (ii) by accounting for beaming and self-heating phenomena in
the YORP computation \citep[e.g.,][]{rg2012,rg2013}. We note that the latter effects
tend to decrease the predicted $\upsilon_{\rm mod}$ value, suitably approaching the
observed value. On the other hand, effects of the lateral heat conduction in
centimeter- to decimeter-sized surface irregularities would produce an additional
acceleration component in $\upsilon_{\rm mod}$ \citep[e.g.,][]{gk2012}. Since
this component is not significantly apparent, we assume the surface of (10115)
1992~SK is not overly rugged. This is also in agreement with conclusions driven
from the analysis of radar data by \citet{Bus.ea:06}, who interpreted the radar
circular polarization measurements as an indication of a similarity to the
surface of Eros.

Our runs also provide the predicted value of the Yarkovsky effect, namely
a rate of secular change in the semimajor axis. For the abovementioned nominal
values of physical parameters, and the rotation state determined in Sec.~\ref{sec:LC_inversion},
we find that $(\dadt)_{\rm mod}$ ranges between $-1.5\times 10^{-4}$ and
$-4\times 10^{-4}$ au~Myr$^{-1}$, with a maximum value for the surface thermal
inertia $\simeq 260$ J~m$^{-1}$~K$^{-1}$~s$^{-1/2}$ \citep[statistically quite
plausible value, e.g.,][]{delbo2015}.
These values are slightly smaller than the value measured for (6489) Golevka 
\citep{cetal2003}, which is the first such case in history, or even larger than the value 
determined for (1685) Toro \citep{Dur.ea:18a}. Yet, no statistically robust 
$\dadt$ drift has been detected from the orbital fit of (10115) 1992~SK so far.
The principal difference with the exemplary cases mentioned above
consists of a significantly poorer dataset of accurate astrometric observations
for (10115) 1992~SK. In particular, this asteroid was radar-sensed
during only one close approach to the Earth in March 1999 \citep{Bus.ea:06},
while Golevka and Toro had radar observations over several approaches
to the Earth, well-separated in time. The availability of radar data still appears to be
a decisive quality for the Yarkovsky effect determination, especially for
kilometer-sized targets \citep[e.g.,][]{vetal2015}.

\subsection{(1685) Toro}
We scaled our shape model for (1685) Toro such that its equivalent volumic size
was $3.5$~km derived by \citet{Dur.ea:18a} from a combination of optical and infrared
photometry. This value is consistent but slightly smaller than a determination by
\citet{netal2016}, who obtained $3.91$~km from NEOWISE observations. We set the
nominal bulk density of $2.5$ g~cm$^{-3}$ and used the rotation state from Sec.~\ref{sec:LC_inversion}.
With those parameters, we obtained a theoretical value of $\upsilon_{\rm mod} = 8.8\times
10^{-9}$ rad~d$^{-2}$ by YORP, independently of the surface thermal inertia. This
value favorably compares with $\upsilon_{\rm mod} = 10.4\times 10^{-9}$ rad~d$^{-2}$,
obtained by \citet{Dur.ea:18a} for a slightly different shape and pole parameters of Toro.
This indicates the stability of the nominal YORP prediction, but also shows that a
realistic uncertainty of the predictions is at the level $\sim 10^{-9}$ rad~d$^{-2}$.
At the same time, $\upsilon_{\rm mod}$ is larger by a factor $\sim 3$  than $\upsilon$
determined from the observations. The discrepancy could be made smaller by adopting the
larger size from \citet{netal2016}, larger bulk density, and shape variations on scales,
which cannot be constrained by observations. Therefore we consider the comparison
between $\upsilon$ and $\upsilon_{\rm mod}$ acceptable, and this justifies our
belief that the detected signal is due to the YORP effect.

We also note that our simulations provided a prediction of the Yarkovsky semimajor
drift $(\dadt)_{\rm mod}$ in the $-0.6\times 10^{-4}$ and $-1.4\times 10^{-4}$ au~Myr$^{-1}$
range, depending on the surface inertia value \citep[see also Fig.~9 in][]{Dur.ea:18a}.
These values match  $\dadt = -(1.39\pm 0.31)\times 10^{-4}$ au~Myr$^{-1}$
very well, which was determined from the astrometric observations of Toro available to date. 

\subsection{(1620) Geographos}
\citet{Dur.ea:08b} reported a robust YORP detection for Geographos with $\upsilon=(1.15\pm
0.15)\times 10^{-8}$ rad~d$^{-2}$ and they also argued that it matches
the theoretically predicted value $\upsilon_{\rm mod} = 1.4\times 10^{-8}$ rad~d$^{-2}$
very
well from their model. Our new value $\upsilon=(1.14\pm 0.03)\times 10^{-8}$ rad~d$^{-2}$,
derived from the available photometric dataset to date, basically confirms the 2008 value
and improves its statistical significance by shrinking the formal uncertainty. Since
the pole direction and shape models are also very similar to those in 2008, we do not 
expect much difference in the theoretically predicted value $\upsilon_{\rm mod}$ either.
Interestingly, scaling our new shape model to $2.56$~km size \citep{Hud.Ost:99} and using a bulk density of
$2.5$ g~cm$^{-3}$, the same values as in 2008, we obtained $\upsilon_{\rm mod} = 0.96\times 
10^{-8}$ rad~d$^{-2}$. A comparison of the two predictions indicates that the Geographos
spin-state and shape configurations provide a little less stable platform for the YORP
predictions. Still, the comparison with the observed $\upsilon$ value is fairly
satisfactory, and there is little doubt about the interpretation of the detected signal.

It is interesting to note that the predicted values of the Yarkovsky semimajor
axis drift for Geographos are very similar to those of Toro mentioned above
\citep[see also Fig.~6 in][]{vetal2005}. The larger size of Toro is apparently
compensated for by several factors: (i) slightly larger obliquity, (ii) lower perihelion,
and (iii) more elongated shape of Geographos, which makes the Yarkovsky effect
smaller \citep[e.g.,][]{v1998b}. In spite of a robust Yarkovsky detection for Toro,
the current astrometry of Geographos permits for a statistically less significant detection of 
$(\dadt) = -(1.33 \pm 0.42) \times 10^{-4}$\,au\,Myr$^{-1}$.
The principal reason consists of a wealth of radar data for Toro, suitably distributed over
four close encounters with the Earth (including accurate measurements in 2016),
and a poorer set of radar measurements for Geographos, over just two close
encounters to the Earth in 1983 and 1994. \citet{vetal2005} were expecting that the
Yarkovsky effect would be firmly detected in the orbit of Geographos by now, but the key
element they assumed were accurate astrometric observations in 2008 and/or 2019. Radar
observations during the close approach in August 2026 might be an alternative option
unless the distance is too large for existing radar systems.

\section{Conclusions}

It is interesting to compare our determined value of $\upsilon = \domdt$
for (10115) 1992~SK with two other exemplary asteroids with good detection of
the YORP effect. We ask the readers to first consider the case of (101955) Bennu, for which 
\citet{hetal2019} obtained $\upsilon = (6.34\pm 0.91)\times 10^{-8}$ rad~d$^{-2}$,
which is slightly smaller than $\upsilon = (8.3\pm 0.6)\times 10^{-8}$ rad~d$^{-2}$ of
(10115) 1992~SK. The two asteroids have approximately the same heliocentric
orbits, whose difference has only a very small impact on the YORP effect strength,
but the principal difference is due to the following: (i) about twice as small of a size of Bennu as opposed to
1992~SK, (ii) about twice as small of a density of Bennu as opposed to 1992~SK, and (iii) a larger
obliquity of Bennu as opposed to 1992~SK \citep[see, e.g.,][]{letal2019}. In a simple YORP
approach, refraining from the detailed influence of the asteroid shape at various
scales, one would thus favor the YORP strength on Bennu by nearly an order of magnitude
over that on 1992~SK. Yet, the YORP effect is some $25$\% smaller for Bennu.
This clearly demonstrates that the shape of Bennu is relatively unfavorable to
a strong YORP effect, which is probably due to its high degree of rotational symmetry.
On the contrary, the YORP-induced acceleration of (1862)~Apollo is
$\upsilon = (5.5\pm 1.2)\times 10^{-8}$ rad~d$^{-2}$ \citep[e.g.,][]{ketal2007,Dur.ea:08},
which is only a factor of $1.5$ smaller than for 1992~SK. The two asteroids again have similar
orbits, both are S-type objects, such that we do not have an a priori reason to
suspect a very different bulk density, and even their obliquity is also similar.
The main difference then is in about a $50$\% larger size of Apollo, such that
naively we would expect a factor of $\simeq 2.25$ in their YORP strength (favoring
1992~SK). This is not very different from the observed factor of $\simeq 1.5$,
implying that Apollo and 1992~SK are similarly favorable to a non-negligible
YORP strength. Indeed, their large-scale resolved shape models are somewhat
similar and lack a high degree of symmetry. These examples clearly demonstrate
the well known high significance of details of the shape model for the strength of
the YORP effect \citep[see][and references therein]{vetal2015}.

A robust result for Geographos, and a slightly weaker, but plausible one, for Toro, illustrate
that the possibility to detect the YORP effect is not reserved to the category of
very small near-Earth asteroids. Instead, the $2$ to $4$~km class of asteroids is fully
accessible for YORP detections if good data are spread over an amenable time interval
of a few decades \citep[see also][for a more formal analysis]{rg2013det}. In this
respect, it might be interesting to carefully review available data for a few kilometer-sized
objects and re-analyze their early observations in the 1950s or 1960s. While the astrometric
information was used from these frames, the value for photometry had not been tested yet. It is
possible that some of these data may reveal interesting constraints on YORP if
properly analyzed.

It is also interesting to overview YORP detections that have been achieved
so far. Five pre-2015 cases have been summarized in the review chapter by
\citet{vetal2015}. Since then, the YORP detection has been reported for (161989)
Cacus by \citet{Dur.ea:18a}, (101955) Bennu by \citet{hetal2019}, (68346) 2001~KZ66 by \cite{Zeg.ea:21}, and
\citet{retal2019} discussed a plausible YORP determination in the case of (85990)
1999~JV6 (though here its significance is only marginal because of a still short
arclength covered by the observations). In this paper, we added a robust
YORP detection for (10115) 1992~SK and argue for a weak, but very plausible YORP signal
in the case of (1685) Toro \citep[see also][]{Dur.ea:18a}. Amazingly enough, all these cases
have $\upsilon$ positive, thus implying acceleration of the rotation rate.

The simplest variants of the YORP effect modeling, starting with \citet{rub2000},
\citet{vc2002}, and \citet{Cap.Vok:04}, predict about an equal likelihood of
a positive and negative value for $\upsilon$ (i.e., rotational acceleration or
deceleration by YORP). This result does not appear to change when effects
of thermal beaming, due to unresolved small-scale irregularities, and
self-irradiation of surface elements are added to the computation, though
the overall magnitude of $\upsilon$ may be decreased \citep[e.g.,][]{rg2012,rg2013}.
In all these approaches, the thermal modeling is restricted to a one-dimensional
conduction below a particular surface facet. Thermal communication of different
facets is neglected or limited to mutual irradiation in the model of \citet{rg2013},
but no thermal communication of the surface units is allowed by internal conduction.
The idea of the importance of the thermal communication of surface facets via
conduction on small-scale surface features was discovered by \citet{gk2012},
and further studied by \citet{getal2014}, \citet{setal2015}, or \citet{setal2016}
in more complicated geometries and configurations. The fundamental aspect of
mutual conductive contact of different surface units is that it breaks the
symmetry in predicted positive and negative $\upsilon$ values for a sample
of asteroidal shape models. Instead, the $\upsilon$ value more likely becomes
positive than negative, but the degree of this asymmetry depends on a large
number of parameters that cannot be easily predicted. So eventually, observations
may help to set this asymmetry degree. Given this perspective, we might interpret
the YORP detections achieved thus far -- not counting the result of \cite{retal2019} for (85990) 1999~JV6 -- as evidence for a $> 10:1$ favor in YORP
making the rotation accelerated over decelerated for kilometer-sized near-Earth
objects.

While we do not see any a priori selection bias of the asteroids for which YORP
has been detected so far, it would be interesting to enlarge the sample of asteroids with a YORP detection for those with larger periods. It is obviously dangerous to draw far-reaching conclusions
from a still, very limited sample of only ten objects and even more
dangerous to attempt to extrapolate our $\upsilon$-asymmetry guess to a class of
few- to ten-kilometer-sized objects in the main belt. Yet, it would be interesting
to do so because this population offers some interesting hints about the YORP
influence. In particular, \citet{petal2008}, and more recently \citet{petal2020},
show that there is a significant fraction of these main-belters which rotate slowly.
In the situation when planetary encounters are not effective, the YORP effect is
suspected to be the primary mechanism to explain this slowly rotating subpopulation.
If this is the case, large asymmetry in positive versus negative $\upsilon$ values would
imply that the small fraction of asteroids driven to slow rotations must remain in
this state for a very long time. But this, in turn, would have an interesting
implication on properties of collisional dynamics in the main belt population of
objects. However, it is also possible that kilometer-sized, and smaller, near-Earth
asteroids have surface properties different from order-of-magnitude larger
main belt asteroids. In this case, their $\upsilon$ asymmetry degree may be
different. Because the YORP effect is unlikely to be detected among the main-belt
asteroids in the foreseeable future, a continuing effort to characterize
the YORP properties among a near-Earth population remains a primary objective.
In particular, enlarging the sample of cases with both a YORP detection and YORP
nondetections \citep[such as 1865 Cerberus discussed in][]{Dur.ea:12b} is vital for
constraining theoretical concepts of this interesting phenomenon in planetary science.

  \begin{acknowledgements}
    This work has been supported by the Czech Science Foundation grant 20-04431S. This research has made use of the KMTNet system operated by the Korea Astronomy and Space Science Institute (KASI) and the data were obtained at one of three host sites, SAAO in South Africa. The work at Abastumani was supported by the Shota Rustaveli National Science Foundation, Grant RF-18-1193. The observational data was partially obtained with the 1 m telescope in Simeiz of the Center for collective use of INASAN.
  \end{acknowledgements}

\newcommand{\SortNoop}[1]{}

  \clearpage
  
  \begin{appendix}
  
    \section{Uncertainty of the YORP parameter}
        
                        The uncertainties of parameters derived from light curve inversion presented in the main text were determined with bootstrap (Sect.~\ref{sec:LC_inversion}). To have an independent estimate, we used the same approach as \cite{Vok.ea:11} or \cite{Pol:14} and computed $\chi^2$ for different fixed values of $\upsilon$ (all other parameters were optimized). Then we defined the $1\sigma$ uncertainty interval of $\upsilon$ as such that $\chi^2$ increased by a factor of $1 + \sqrt{2/\nu}$, where $\nu$ is the number of degrees of freedom. These intervals are larger than those determined by bootstrap. 
                        
      \begin{figure}[h]
        \begin{center}
          \includegraphics[width=\columnwidth]{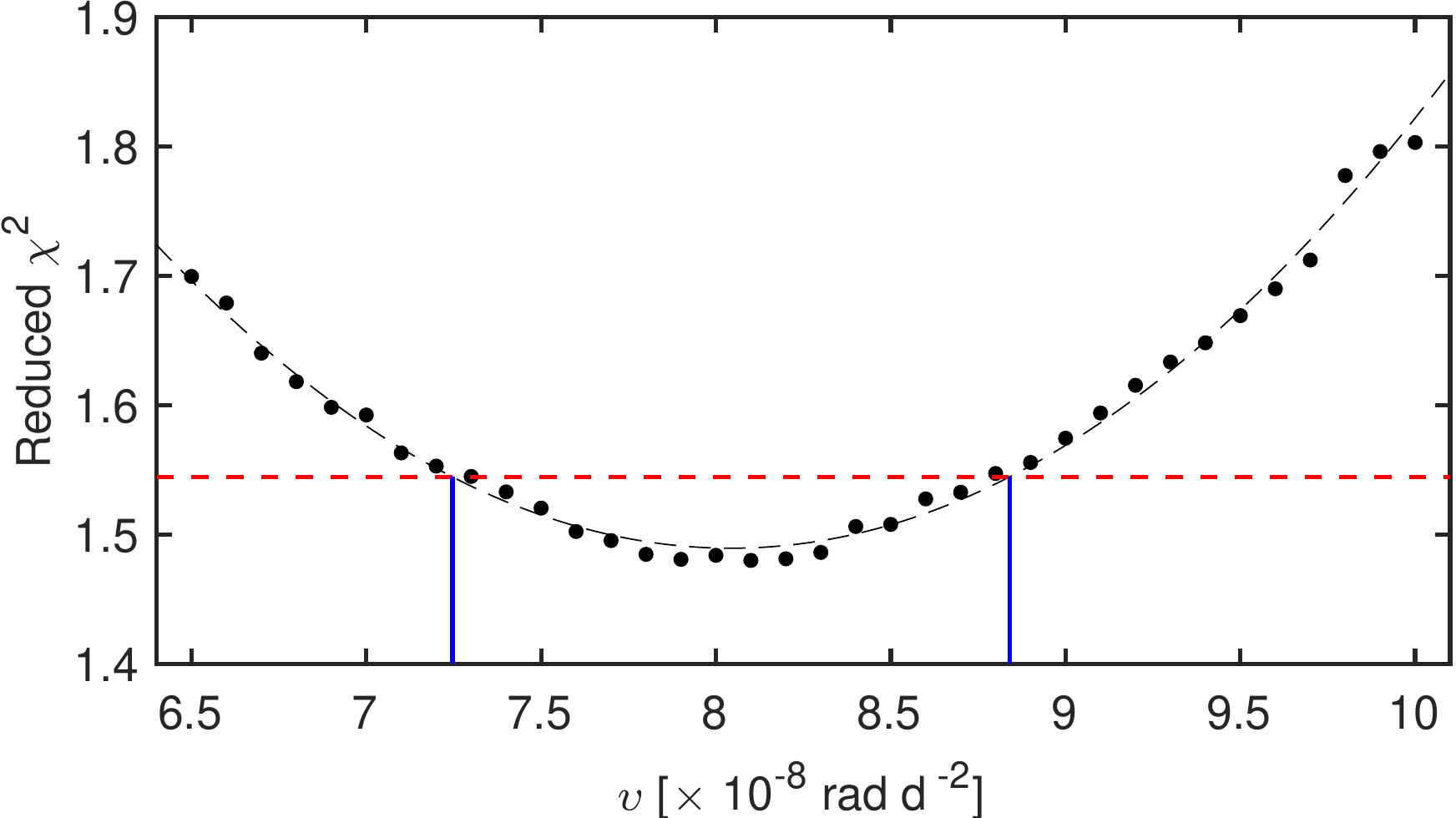}
          \caption{\label{fig_10115_YORP}
           Dependence of the goodness of the fit measured by the reduced $\chi^2$ (defined as $\chi^2/\nu$, where $\nu$ is the number of degrees of freedom) on the YORP parameter $\upsilon$ for asteroid (10115) 1992~SK. The dashed curve is a quadratic fit of the data points. The dashed red line indicates a $3.7\%$ increase in the $\chi^2$, which defines our $1\sigma$ uncertainty interval of $\pm 0.8 \times 10^{-8}\,\text{rad}\,\text{d}^{-2} $ given the number of degrees of freedom.}
        \end{center}
      \end{figure}

      \begin{figure}[h]
        \begin{center}
          \includegraphics[width=\columnwidth]{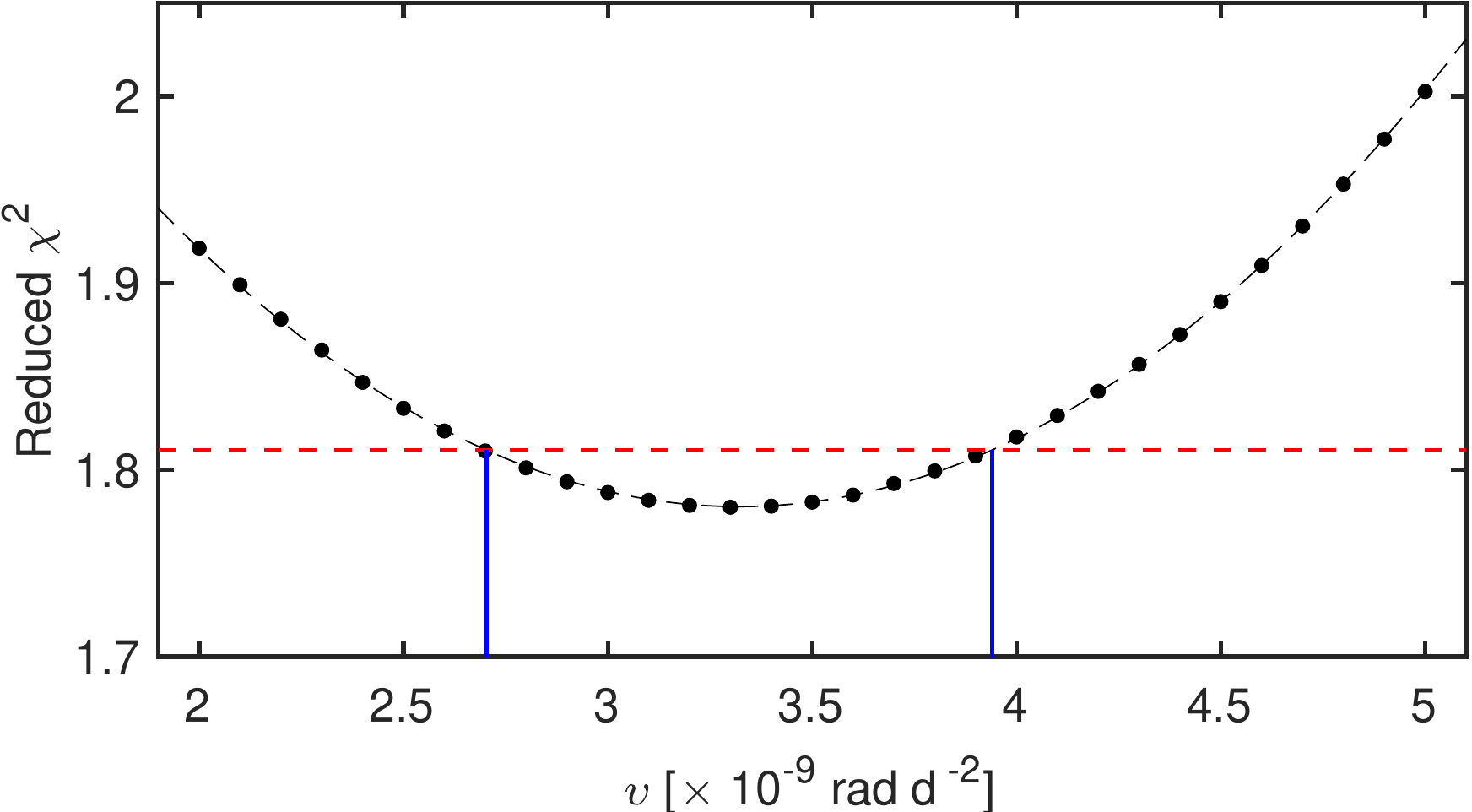}
          \caption{\label{fig_1685_YORP}
           Dependence of the goodness of the fit measured by the reduced $\chi^2$ on the YORP parameter $\upsilon$ for asteroid (1685)~Toro. The dashed curve is a quadratic fit of the data points. The dashed red line indicates a $1.7\%$ increase in the $\chi^2$, which defines our $1\sigma$ uncertainty interval of $\pm 0.6 \times 10^{-9}\,\text{rad}\,\text{d}^{-2} $ given the number of degrees of freedom.}
        \end{center}
      \end{figure}
      
      \begin{figure}[h]
        \begin{center}
          \includegraphics[width=\columnwidth]{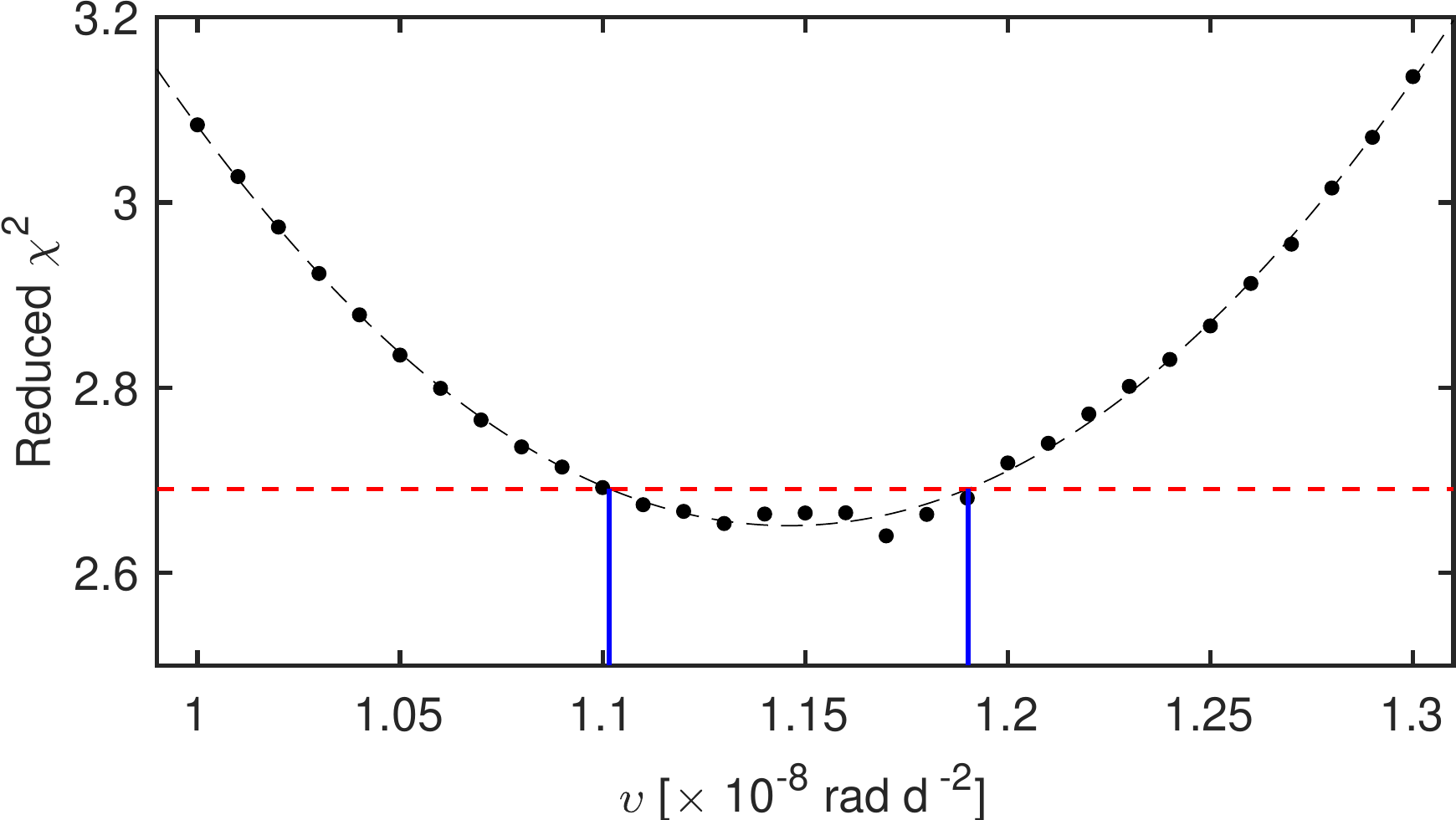}
          \caption{\label{fig_1620_YORP}
           Dependence of the goodness of the fit measured by the reduced $\chi^2$ on the YORP parameter $\upsilon$ for asteroid (1620) Geographos. The dashed curve is a quadratic fit of the data points. The dashed red line indicates a $1.5\%$ increase in the $\chi^2$, which defines our $1\sigma$ uncertainty interval of $\pm 0.04 \times 10^{-8}\,\text{rad}\,\text{d}^{-2} $ given the number of degrees of freedom.}
        \end{center}
      \end{figure}

    \clearpage  
        
    \section{Shape models}  
    
      \begin{figure*}[t]
        \begin{center}
          \includegraphics[width=1.5\columnwidth]{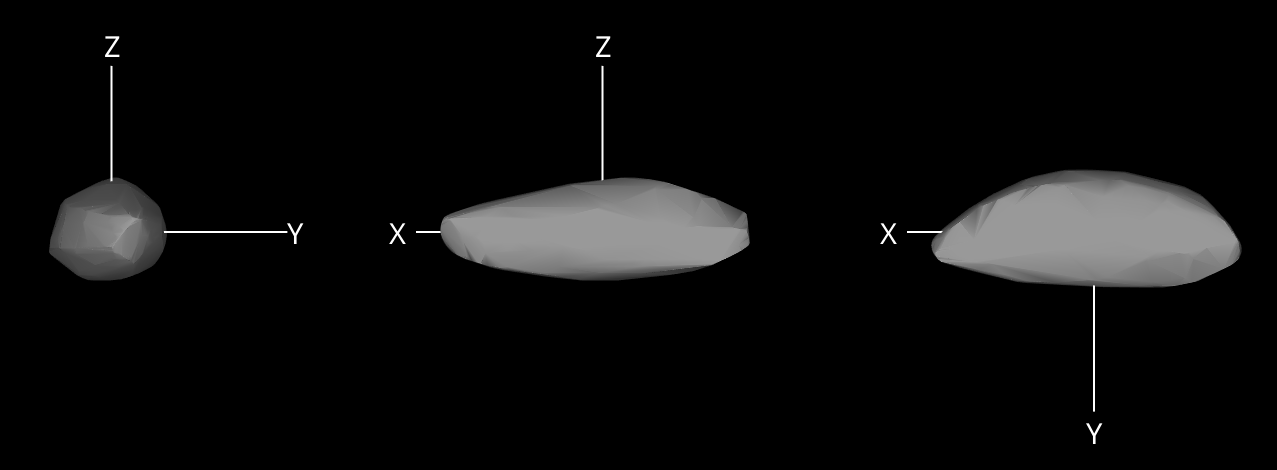}
          \caption{\label{fig_1620_shape}
          Shape model of asteroid (1620)~Geographos shown from equatorial level (left and center, $90\degr$ apart) and pole-on (right).}
        \end{center}
      \end{figure*}

      \begin{figure*}[h]
        \begin{center}
          \includegraphics[width=1.5\columnwidth]{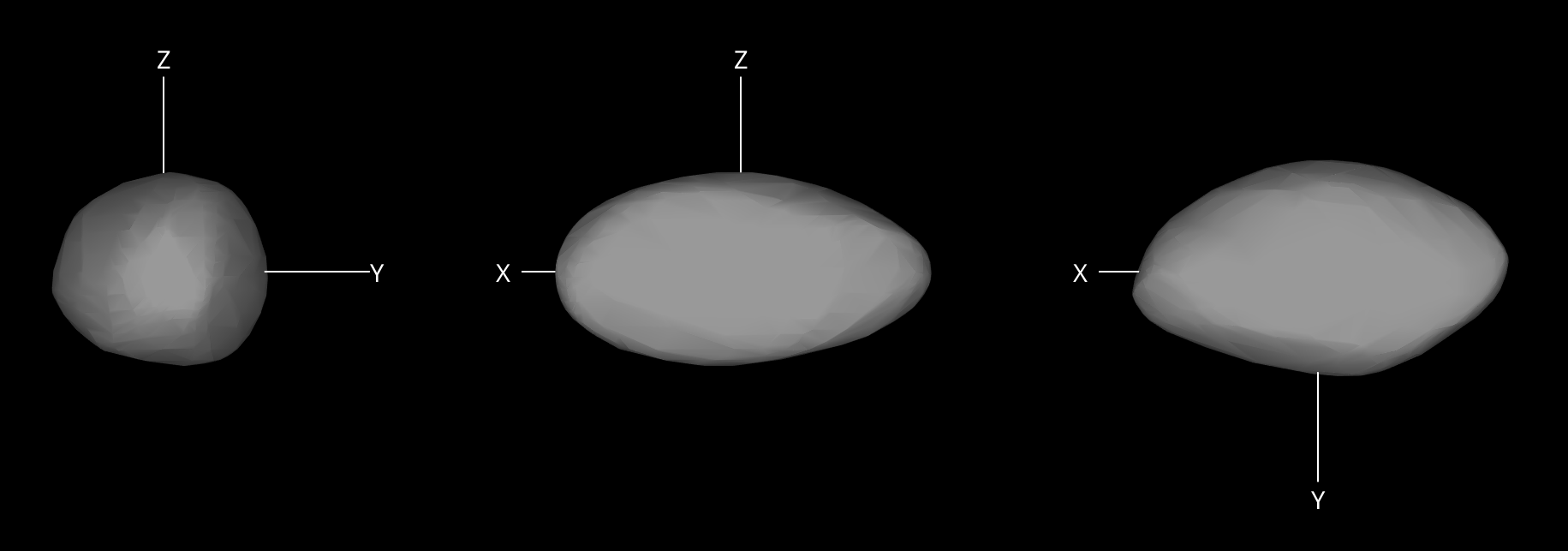}
          \caption{\label{fig_1685_shape}
          Shape model of asteroid (1685)~Toro shown from equatorial level (left and center, $90\degr$ apart) and pole-on (right).}
        \end{center}
      \end{figure*}
      
    \clearpage
  
                \section{New photometric observations}
                
    \begin{table*}[h]
      \caption{\label{tab:aspect_10115} 
      Aspect data for observations of (10115) 1992~SK.}
      \centering
      \begin{tabular}{cccrrrl}
        \hline \hline
        Date    & $r$   & $\Delta$      & $\alpha\phantom{g}$   & \multicolumn{1}{c}{$\lambda$} & \multicolumn{1}{c}{$\beta$}     & Observatory or        \\
        & [au]  & [au]          & [deg]                 & \multicolumn{1}{c}{[deg]}         & \multicolumn{1}{c}{[deg]}     &       Reference \\
        \hline
        1999 02 08.0  & 1.195    & 0.305  & 41.3     & 158.3     & $50.4$ & 1 \\
1999 02 09.1  & 1.189    & 0.298  & 41.6     & 158.1     & $50.9$ & 1 \\
1999 02 09.9  & 1.185    & 0.292  & 41.8     & 157.9     & $51.2$ & 1 \\
1999 02 15.9  & 1.153    & 0.253  & 44.2     & 155.6     & $53.9$ & 1 \\
1999 03 11.9  & 1.028    & 0.109  & 68.8     & 117.2     & $63.5$ & 1 \\
1999 03 13.0  & 1.022    & 0.103  & 71.3     & 112.4     & $63.4$ & 1 \\
2006 03 06.1  & 1.017    & 0.132  & 75.6     & 246.0     & $40.9$ & 2 \\
2006 03 07.0  & 1.012    & 0.128  & 77.7     & 249.7     & $39.7$ & 2 \\
2013 11 03.3  & 1.591    & 0.624  & 12.8     &  34.0     & $19.8$ & 3 \\
2013 11 05.3  & 1.595    & 0.633  & 13.7     &  32.9     & $20.1$ & 3 \\
2013 11 06.3  & 1.597    & 0.637  & 14.1     &  32.3     & $20.2$ & 3 \\
2017 09 26.9  & 1.193    & 0.426  & 53.9     & 290.7     & $-18.2$ & KMTNet-SAAO \\
2017 10 14.1  & 1.279    & 0.562  & 48.2     & 307.7     & $-8.0$ & DK \\
2017 10 14.2  & 1.280    & 0.563  & 48.2     & 307.8     & $-7.9$ & DK \\
2017 10 19.1  & 1.303    & 0.609  & 47.3     & 311.8     & $-5.8$ & DK \\
2017 10 19.2  & 1.304    & 0.610  & 47.2     & 311.9     & $-5.7$ & DK \\
2017 11 12.1  & 1.409    & 0.867  & 44.0     & 328.2     & $ 1.4$ & DK \\
2020 01 12.3  & 1.200    & 0.635  & 54.9     & 197.9     & $22.0$ & DK \\
2020 10 17.8  & 1.575    & 0.644  & 20.2     &  53.3     & $17.3$ & BOAO \\
2020 10 18.8  & 1.578    & 0.642  & 19.6     &  52.8     & $17.6$ & BOAO \\
2020 10 19.3  & 1.579    & 0.642  & 19.2     &  52.5     & $17.8$ & BOAO \\
2020 10 19.8  & 1.580    & 0.641  & 18.9     &  52.3     & $18.0$ & LOAO \\
2020 10 20.3  & 1.581    & 0.640  & 18.6     &  52.0     & $18.1$ & LOAO \\
2020 10 21.3  & 1.584    & 0.639  & 18.0     &  51.5     & $18.4$ & LOAO \\
2020 11 17.0  & 1.630    & 0.695  & 17.4     &  36.2     & $23.3$ & Rozhen \\
2020 12 04.1  & 1.647    & 0.808  & 25.9     &  30.0     & $22.9$ & DK \\
2020 12 12.8  & 1.652    & 0.881  & 29.4     &  28.7     & $22.2$ & Nauchny \\

        \hline
      \end{tabular}
      \tablefoot{The table lists asteroid's distance from the Sun $r$ and from the Earth $\Delta$, the solar phase angle $\alpha$, the geocentric ecliptic coordinates of the asteroid $(\lambda, \beta)$, and the observatory or source (KMTNet-SAAO -- Korea Microlensing Telescope Network-South African Astronomical Observatory, 1.6\,m; BOAO -- Bohyunsan Optical Astronomy Observatory, 1.8\,m; LOAO -- Lemmonsan Optical Astronomy Observatory, 1\,m; DK -- Danish telescope, La Silla, 1.54\,m; Rozhen -- Rozhen Observatory, 2\,m; Nauchny -- Crimean Astrophysical Observatoty, 2.6\,m).}
      \tablebib{(1) \cite{Bus.ea:06}; (2) \cite{Pol:12}; (3) \cite{War:14e}}
    \end{table*}

    \begin{table*}[h]
      \caption{\label{tab:aspect_1620}  
        Aspect data for new observations of (1620)~Geographos.}
        \centering
        \begin{tabular}{cccrrrl}
            \hline \hline
            Date        & $r$   & $\Delta$      & $\alpha\phantom{g}$   & \multicolumn{1}{c}{$\lambda$}   & \multicolumn{1}{c}{$\beta$}   & Observatory   or \\
            & [au]      & [au]          & [deg]                 & \multicolumn{1}{c}{[deg]}         & \multicolumn{1}{c}{[deg]}     &       Reference \\
            \hline
            2008 09 01.0  & 1.263    & 0.587  & 51.6     &  57.2     & $13.7$ & Simeiz \\
2008 09 04.0  & 1.278    & 0.581  & 50.1     &  57.5     & $14.9$ & Simeiz \\
2008 10 27.8  & 1.506    & 0.568  & 20.4     &  38.1     & $31.7$ & 1 \\
2008 10 28.8  & 1.509    & 0.571  & 20.3     &  37.5     & $31.7$ & 1 \\
2008 11 04.3  & 1.530    & 0.598  & 20.5     &  33.5     & $31.6$ & 2 \\
2008 11 06.3  & 1.536    & 0.608  & 20.8     &  32.4     & $31.4$ & 2 \\
2008 11 07.3  & 1.539    & 0.613  & 21.0     &  31.8     & $31.3$ & 2 \\
2008 11 18.2  & 1.569    & 0.681  & 24.2     &  27.0     & $29.7$ & 2 \\
2008 11 19.2  & 1.572    & 0.688  & 24.5     &  26.7     & $29.5$ & 2 \\
2008 11 20.3  & 1.574    & 0.696  & 24.9     &  26.4     & $29.3$ & 2 \\
2011 12 29.2  & 1.622    & 0.697  & 18.1     &  86.5     & $29.1$ & App \\
2011 12 29.8  & 1.621    & 0.697  & 18.2     &  86.2     & $29.0$ & AAO \\
2012 10 13.7  & 1.160    & 0.563  & 59.3     & 292.3     & $ 7.0$ & AAO \\
2012 10 14.7  & 1.165    & 0.570  & 58.8     & 293.5     & $ 7.3$ & AAO \\
2012 10 16.8  & 1.176    & 0.586  & 57.8     & 296.0     & $ 7.9$ & Simeiz \\
2012 10 17.7  & 1.182    & 0.594  & 57.4     & 297.1     & $ 8.1$ & Simeiz \\
2015 07 30.9  & 1.185    & 0.911  & 56.1     &  51.5     & $ 5.5$ & Tien-Shan \\
2015 08 05.9  & 1.217    & 0.904  & 54.8     &  54.5     & $ 6.9$ & Tien-Shan \\
2015 08 18.9  & 1.284    & 0.877  & 51.8     &  60.2     & $10.1$ & Tien-Shan \\
2015 08 19.9  & 1.289    & 0.875  & 51.5     &  60.6     & $10.4$ & Tien-Shan \\
2015 09 22.0  & 1.437    & 0.759  & 41.5     &  68.8     & $19.7$ & AAO \\
2015 09 22.9  & 1.441    & 0.756  & 41.1     &  68.9     & $20.0$ & AAO \\
2015 12 13.2  & 1.647    & 0.824  & 26.9     &  38.1     & $27.3$ & 3 \\
2015 12 15.2  & 1.649    & 0.840  & 27.6     &  37.7     & $26.9$ & 3 \\
2015 12 16.2  & 1.650    & 0.848  & 27.9     &  37.6     & $26.6$ & 3 \\
2019 01 03.3  & 1.571    & 0.633  & 17.1     & 103.6     & $28.0$ & 4 \\
2019 01 04.3  & 1.568    & 0.630  & 17.1     & 102.9     & $27.9$ & 4 \\
2019 01 10.2  & 1.552    & 0.617  & 17.9     &  99.2     & $27.3$ & 4 \\
2019 01 11.3  & 1.549    & 0.615  & 18.2     &  98.5     & $27.1$ & 4 \\
2019 01 21.0  & 1.520    & 0.612  & 22.7     &  92.8     & $25.0$ & AAO \\
2019 03 06.8  & 1.343    & 0.746  & 46.7     &  85.9     & $10.1$ & Simeiz \\
2019 09 24.3  & 1.150    & 0.248  & 48.6     & 302.6     & $14.6$ & 5 \\
2019 09 28.2  & 1.172    & 0.279  & 47.0     & 307.2     & $16.1$ & 5 \\
2019 09 29.2  & 1.177    & 0.287  & 46.7     & 308.3     & $16.4$ & 5 \\
2019 09 30.2  & 1.182    & 0.295  & 46.4     & 309.4     & $16.7$ & 5 \\
2019 10 12.8  & 1.249    & 0.406  & 44.1     & 320.1     & $18.7$ & Simeiz \\
2019 11 06.5  & 1.368    & 0.655  & 42.6     & 335.5     & $18.8$ & BMO \\
2019 11 07.5  & 1.373    & 0.666  & 42.6     & 336.0     & $18.7$ & BMO \\
2019 11 11.5  & 1.391    & 0.709  & 42.4     & 338.3     & $18.5$ & BMO \\
2019 11 14.4  & 1.403    & 0.742  & 42.3     & 339.9     & $18.4$ & BMO \\
2019 11 15.5  & 1.408    & 0.753  & 42.2     & 340.4     & $18.3$ & BMO \\
2019 11 18.5  & 1.420    & 0.787  & 42.1     & 342.0     & $18.1$ & BMO \\
2019 11 19.5  & 1.424    & 0.798  & 42.0     & 342.6     & $18.0$ & BMO \\
2019 11 20.4  & 1.428    & 0.809  & 42.0     & 343.1     & $18.0$ & BMO \\

        \hline
        \end{tabular}
        \tablefoot{The table lists asteroid's distance from the Sun $r$ and from the Earth $\Delta$, the solar phase angle $\alpha$, the geocentric ecliptic coordinates of the asteroid $(\lambda, \beta)$, and the observatory or source (AAO -- E.Kharadze Abastumani Astrophysical Observatory, 70\,cm; BMO -- Blue Mountains Observatory, 35\,cm; Simeiz -- Simeiz Observatory, 1\,m; Tien-Shan -- Tien-Shan Astronomical Observatory, 1\,m; App -- Appalachian State University's Dark Sky Observatory, 81\,cm.)}
        \tablebib{(1) \cite{Pol:09}; (2) \cite{Ski.ea:19}; (3) \cite{War:16e}; (4) \cite{War.Ste:19m}; (5) \cite{War.Ste:20a}}
    \end{table*}

                \begin{table*}[h]
      \caption{\label{tab:aspect_1685}  
        Aspect data for new observations of (1685)~Toro.}
        \centering
        \begin{tabular}{cccrrrl}
            \hline \hline
            Date        & $r$   & $\Delta$      & $\alpha\phantom{g}$   & \multicolumn{1}{c}{$\lambda$}   & \multicolumn{1}{c}{$\beta$}   & Observatory or      \\
            & [au]      & [au]          & [deg]                 & \multicolumn{1}{c}{[deg]}         & \multicolumn{1}{c}{[deg]}     &       Reference \\
            \hline
            2018 05 20.9  & 1.956    & 1.001  & 13.8     & 215.5     & $-13.4$ & Wise  \\
2018 06 07.8  & 1.937    & 1.102  & 22.8     & 210.1     & $-11.1$ & Wise \\
2018 06 08.8  & 1.936    & 1.110  & 23.2     & 209.9     & $-10.9$ & Wise  \\
2018 06 09.8  & 1.935    & 1.117  & 23.6     & 209.7     & $-10.8$ & Wise  \\
2018 06 22.8  & 1.914    & 1.223  & 28.2     & 208.6     & $-9.0$ & Wise  \\
2020 06 15.4  & 1.317    & 0.593  & 47.6     & 337.5     & $ 5.7$ &   1 \\
2020 06 17.4  & 1.304    & 0.572  & 48.2     & 339.3     & $ 6.2$ &   1 \\
2020 06 18.4  & 1.297    & 0.562  & 48.5     & 340.3     & $ 6.5$ &   1 \\
2020 06 19.4  & 1.290    & 0.552  & 48.9     & 341.2     & $ 6.9$ &   1 \\
2020 06 21.4  & 1.277    & 0.532  & 49.6     & 343.2     & $ 7.5$ &   1 \\
2020 06 22.4  & 1.270    & 0.523  & 50.0     & 344.3     & $ 7.8$ &   1 \\
2020 06 23.4  & 1.263    & 0.513  & 50.5     & 345.3     & $ 8.2$ &   1 \\
2020 06 27.4  & 1.236    & 0.477  & 52.3     & 349.8     & $ 9.7$ &   1 \\
2020 06 29.0  & 1.225    & 0.464  & 53.2     & 351.7     & $10.3$ & Wise \\
2020 06 31.0  & 1.211    & 0.448  & 54.3     & 354.3     & $11.1$ & Wise \\
2020 07 03.0  & 1.197    & 0.432  & 55.5     & 357.0     & $12.0$ & Chuguev \\
2020 07 11.0  & 1.140    & 0.379  & 61.7     &   9.3     & $15.7$ & Chuguev \\
2020 07 13.9  & 1.119    & 0.364  & 64.4     &  14.6     & $17.1$ & Wise \\
2020 07 22.9  & 1.055    & 0.335  & 74.1     &  33.1     & $20.8$ & Kitab \\
2020 07 23.0  & 1.055    & 0.335  & 74.1     &  33.1     & $20.8$ & Kitab \\
2020 07 26.0  & 1.034    & 0.332  & 77.6     &  39.9     & $21.7$ & Wise \\
2020 07 28.9  & 1.013    & 0.331  & 80.9     &  46.5     & $22.3$ & Kitab \\
2020 07 29.9  & 1.006    & 0.332  & 82.0     &  48.8     & $22.4$ & Kitab \\
2020 08 08.0  & 0.945    & 0.354  & 90.9     &  68.7     & $22.3$ & AAO \\
2020 08 09.0  & 0.938    & 0.358  & 91.6     &  70.8     & $22.1$ & AAO \\
2020 08 09.9  & 0.932    & 0.362  & 92.3     &  72.6     & $21.9$ & Kitab \\
2020 08 10.0  & 0.932    & 0.362  & 92.4     &  72.7     & $21.9$ & Wise \\
2020 08 10.9  & 0.926    & 0.366  & 93.0     &  74.6     & $21.7$ & Kitab \\
2020 08 10.9  & 0.926    & 0.366  & 93.0     &  74.6     & $21.7$ & Kitab \\
2020 08 12.0  & 0.919    & 0.372  & 93.7     &  76.7     & $21.5$ & Wise \\
2020 08 19.0  & 0.877    & 0.412  & 96.8     &  88.9     & $19.5$ & AAO \\
2020 08 21.9  & 0.861    & 0.432  & 97.4     &  93.5     & $18.6$ & Kitab \\
2020 08 22.9  & 0.856    & 0.439  & 97.5     &  94.9     & $18.2$ & Kitab \\
2020 09 15.0  & 0.776    & 0.639  & 90.1     & 122.7     & $10.2$ & AAO \\
2021 04 30.1  & 1.878    & 0.991  & 20.1     & 183.1     & $-17.1$ & DK \\
2021 04 30.0  & 1.878    & 0.990  & 20.1     & 183.1     & $-17.2$ & DK \\

        \hline
        \end{tabular}
        \tablefoot{The table lists asteroid's distance from the Sun $r$ and from the Earth $\Delta$, the solar phase angle $\alpha$, the geocentric ecliptic coordinates of the asteroid $(\lambda, \beta)$, and the observatory or source (Wise - Wise Observatory, 75\,cm, see \cite{Pol.Bro:09} for observation and reduction details; DK -- Danish telescope, La Silla, 1.54\,m; Chuguev -- Chuguev Observatory, 70\,cm; Kitab -- Kitab Observatory, 36\,cm; AAO -- E.Kharadze Abastumani Astrophysical Observatoryo, 70\,cm).}
        \tablebib{(1) \cite{War.Ste:20g}.}
    \end{table*}
      
  \end{appendix}

\end{document}